\documentclass[aps,prl,twocolumn,superscriptaddress,amssymb,amsmath]{revtex4}
\usepackage{subfigure,amsmath,bbm}

\newcommand {\ie}{{\it i.e.}}
\newcommand {\eg}{{\it e.g.}}
\newcommand {\Tc}{T_{\mathrm{c}}}
\newcommand {\Hc}{H_{\mathrm{c}}}
\newcommand {\Hin}{H_{\mathrm{in}}}
\newcommand {\Hout}{H_{\mathrm{out}}}
\newcommand {\Jc}{J_{\mathrm{c}}}
\newcommand {\Rn}{R_{yx}^{(n)}}
\newcommand {\Rnf}{R_{yx}^{(nf)}}
\newcommand {\RA}{R_{\mathrm{A}}}

\usepackage[dvips]{graphicx}
\usepackage{bm}
\usepackage{pifont}
\usepackage{amsmath}
\usepackage{color}

\begin{document}

\title{Strongly pinned skyrmionic bubbles and higher-order nonlinear Hall resistances at the interface of Pt/FeSi bilayer}

\author{T. Hori}
\affiliation{Institute of Industrial Science, University of Tokyo, Tokyo 153-8505, Japan}
\affiliation{Department of Applied Physics, University of Tokyo, Tokyo 113-8656, Japan}
\author{N. Kanazawa\footnote{To whom correspondence should be addressed: naoya-k@iis.u-tokyo.ac.jp}}
\affiliation{Institute of Industrial Science, University of Tokyo, Tokyo 153-8505, Japan}
\author{K. Matsuura}
\affiliation{RIKEN Center for Emergent Matter Science (CEMS), Wako, 351-0198, Japan}
\affiliation{Department of Physics, Tokyo Institute of Technology, Tokyo 152-8550, Japan}
\author{H. Ishizuka}
\affiliation{Department of Physics, Tokyo Institute of Technology, Tokyo 152-8550, Japan}
\author{K. Fujiwara}
\affiliation{Institute for Materials Research, Tohoku University, Sendai 980-8577, Japan}
\author{A. Tsukazaki}
\affiliation{Institute for Materials Research, Tohoku University, Sendai 980-8577, Japan}
\author{M. Ichikawa}
\affiliation{Department of Applied Physics, University of Tokyo, Tokyo 113-8656, Japan}
\author{M. Kawasaki}
\affiliation{Department of Applied Physics, University of Tokyo, Tokyo 113-8656, Japan}
\affiliation{RIKEN Center for Emergent Matter Science (CEMS), Wako, 351-0198, Japan}
\author{F. Kagawa}
\affiliation{RIKEN Center for Emergent Matter Science (CEMS), Wako, 351-0198, Japan}
\affiliation{Department of Physics, Tokyo Institute of Technology, Tokyo 152-8550, Japan}
\author{M. Hirayama}
\affiliation{Department of Applied Physics, University of Tokyo, Tokyo 113-8656, Japan}
\affiliation{RIKEN Center for Emergent Matter Science (CEMS), Wako, 351-0198, Japan}
\author{Y. Tokura}
\affiliation{Department of Applied Physics, University of Tokyo, Tokyo 113-8656, Japan}
\affiliation{RIKEN Center for Emergent Matter Science (CEMS), Wako, 351-0198, Japan}
\affiliation{Tokyo College, University of Tokyo, Tokyo 113-8656, Japan}
\date{\today}

\begin{abstract}
Engineering of magnetic heterostructures for spintronic applications has entered a new phase, driven by the recent discoveries of topological materials and exfoliated van der Waals materials. Their low-dimensional properties can be dramatically modulated in designer heterostructures via proximity effects from adjacent materials, thus enabling the realization of diverse quantum states and functionalities. Here we investigate spin-orbit coupling (SOC) proximity effects of Pt on the recently discovered quasi-two-dimensional ferromagnetic state at FeSi surface. Skyrmionic bubbles (SkBs) are formed as a result of the enhanced interfacial Dzyloshinskii-Moriya interaction. The strong pinning effects on the SkBs are evidenced from the significant dispersion in size and shape of the SkBs and are further identified as a greatly enhanced threshold current density required for depinning of the SkBs. The robust integrity of the SkB assembly leads to the emergence of higher-order nonlinear Hall effects in the high current density regime, which originate from nontrivial Hall effects due to the noncollinearity of the spin texture, as well as from the current-induced magnetization dynamics via the augmented spin-orbit torque.
\end{abstract}
\maketitle

\section{Introduction}
Atomically thin magnets and their heterostructures offer an excellent platform for realizing a variety of spin states and spintronic functionality~\cite{Fert,Grunberg,Parkin,Cortie}. 
The recent energetic exploration of magnetic or magnetically-proximitized topological materials~\cite{Tokura,Bernevig,Smejkal,Hasan} and van der Waals materials~\cite{Geim,Gong,Gibertini,Mak,Xu} pushes the frontier as highlighted in the realization of quantum anomalous Hall effect~\cite{Chang,Deng,Serlin,Park} and the advances in valleytronics applications~\cite{Zhong,Seyler}. Simply because a small number of carriers and spins are involved in the ultrathin films, the magnetic states are responsive to external stimuli such as electric currents and gate voltages or susceptible to extrinsic effects such as defects and surface roughness. This enables us to control their spintronic functionalities in various ways, while dramatically reducing the energy consumption.

In addition to the low dimensionality, structural symmetry breaking is particularly important for emergent phenomena induced by spin-orbit coupling (SOC) in the ultrathin films~\cite{Manchon,Panagopoulos,Hellman,Bihlmayer}. As a consequence of the broken inversion symmentry at the surfaces or interfaces, antisymmetric SOC effects arise, such as Rashba effect~\cite{Rashba} and Dzyaloshinskii-Moriya (DM) interaction~\cite{Dzyaloshinsky,Moriya}. Their antisymmetric nature leads to the formation of rich spin textures in momentum- and real-spaces, {\it i.e.}, spin-momentum locking in energy bands~\cite{Manchon,Bihlmayer} and N\'{e}el-type domain walls/skyrmions~\cite{Wiesendanger,Fert2,Bogdanov}. The interplay between the conduction electrons and those spin textures causes various transport phenomena, to name a few, Edelstein effect~\cite{Edelstein}, spin-orbit torques (SOTs)~\cite{Garello,Manchon2}, nonreciprocal transports of the second order in electric field $E$~\cite{Yokouchi,Tokura2}, topological Hall effect~\cite{Neubauer,Lee}, and skyrmion drive~\cite{Jonietz,Woo}.

Recently discovered ferromagnetic surface of the chiral-lattice FeSi constitutes a new form of quasi-two-dimensional (2D) magnetic state~\cite{Ohtsuka,Hori}. While the bulk interior of FeSi is of a nonmagnetic insulating state~\cite{Mattheiss}, multiple experimental techniques have revealed that its surface exhibits both conductive (metallic) behaviors~\cite{Ohtsuka,Hori,Maple,Barth,Paglione} and ferromagnetic ordering~\cite{Ohtsuka,Hori}. Here the ferromagnetic order is confined within a depth of $\sim 3.5$  \AA\ from the surface, which corresponds approximately to the top three surface-Fe layers~\cite{Ohtsuka}. Polar distribution of surface electronic orbitals, which is characterized by quantized Zak phase~\cite{Zak} or topology of electric polarization~\cite{Resta,King,Vanderbilt}, boosts the potential gradient perpendicular to the surface and results in large Rashba spin splitting ($\sim 35$ meV)~\cite{Ohtsuka} despite the relatively low atomic numbers of Fe and Si. Owing to the coexistence of the ferromagnetic-metal properties and the large Rashba SOC at the surface, SOT-induced magnetization switching is realized even at room temperature without external assist magnetic fields~\cite{Hori}. Here we note that there are other arguments for topological aspects of the FeSi surface state in terms of topological Kondo insulators~\cite{Maple} or Weyl semimetals~\cite{Changdar}, partly because the origin of the bulk bandgap still remains an interesting question~\cite{Aeppli,Doniach,Anisimov,Kotliar}.  

In this Article, we explore spintronic properties at the FeSi surface by fabricating an interface with Pt and introducing proximity effects of the intrinsically strong SOC. As in the case of multilayers composed of magnetic materials and heavy elements~\cite{Wiesendanger,Fert2,Everschor}, DM interaction is activated at the Pt/FeSi interface [Fig.~1(a)] and skyrmionic bubbles (SkBs) are formed in an in-plane magnetization background. The SkBs vary considerably in size and shape due to inherent structural disorders such as defects, surface roughness, grain boundaries and so on. Also, their motions are highly resistant to the electric current as a result of the strong pinning from those structural disorders. The application of high current density causes nonlinear responses of the SkBs without destroying them or making them under steady flow, leading to higher-order harmonics of Hall signals with respect to the fundamental frequency $f$ of an input ac current.

\begin{figure}
\begin{center}
\includegraphics*[width=8.5cm]{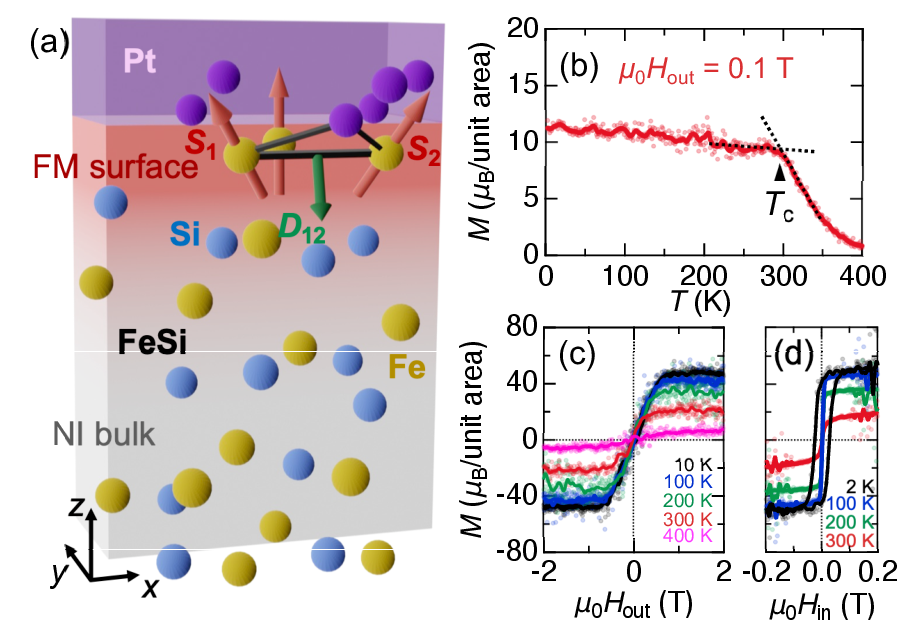}
\caption{
Magnetic properties at the interface between Pt ($t=3$~nm) and FeSi ($t=5$~nm). (a) Schematic picture of the Pt/FeSi bilayer. As a consequence of the spin-orbit coupling (SOC) proximity of Pt, the interfacial Dzyaloshinskii-Moriya (DM) interaction twists the spins at ferromagnetic-metal (FM) surface of FeSi with the nonmagnetic-insulating (NI) bulk, inducing noncollinear spin textures. Here $\bm{D}_{12}$ represents the DM vector acting on spins $\bm{S}_1$ and $\bm{S}_2$. (b) Temperature dependence of magnetization $M$ under an out-of-plane magnetic field $\mu_0 \Hout =0.1$~T. The kink around $T=295$ K represents the helimagnetic ordering. (c) and (d) Out-of-plane and in-plane magnetic-field dependence of magnetization at various temperatures. The measured data (color dots) are smoothed (color lines) for clarity. The unit area corresponds to $\sqrt{3}a^2$, where $a$ is the lattice constant of the cubic unit cell of FeSi.
}
\end{center}
\end{figure}

\section{Methods}
The Pt/FeSi bilayer was fabricated on an insulating Si(111) substrate in an {\it in-situ} setup combining molecular beam epitaxy (MBE) and sputtering methods. While the FeSi(111) layer with thickness $t=5$~nm was epitaxially grown on the Si substrate by MBE~\cite{Sirringhaus}, the Pt layer with $t=3$~nm was deposited at room temperature by sputtering. Also see Refs.~\cite{Ohtsuka,Hori} for the detailed growth procedures. The epitaxial growth of FeSi was confirmed by $\theta$-$2\theta$ x-ray diffraction method, while the surface roughness was estimated by atomic force microscopy (AFM) [Supplemental Material~\cite{supple}, Fig.~S1].
The thin-film sample was processed into Hall-bar devices of 30-$\mu$m width and 40-$\mu$m length by using UV lithography and Ar ion milling. The Hall-bar devices were connected to electrodes made of Au(45~nm)/Ti(5~nm) by electron beam deposition.

Magnetization measurements were performed by using the reciprocating sample option of a magnetic property measurement system (MPMS, Quantum Design). Hall resistance measurements were performed by using a lock-in technique (SR-830, Stanford Research Systems), where higher-harmonic Hall voltages up to the seventh order in response to an input of ac current were measured.

Frequency-modulated magnetic force microscopy (MFM) was performed in noncontact mode with a lift height of 160 nm in a commercially available scanning probe microscope (AFM/MFM I, attocube) on a Hall-bar device. We used the MFMR tip (supplied by NANOSENSORS). Electric current pulses were injected into the Hall-bar device before each MFM scan for evaluating the magnetic domain response to the current.

\section{Results and Discussion}
Magnetization ($M$) measurements revealed that the interfacial state shows ferromagnetic behaviors with in-plane anisotropy below the ordering temperature $\Tc\approx 295$~K [Figs.~1(b)-(d)]. Here we convert the magnetization size to a value per unit area of the FeSi(111) surface. (The unit area corresponds to $\sqrt{3}a^2$, where $a$ is the lattice constant of the cubic unit cell of FeSi.) The magnetization size gradually grows with decreasing temperature $T$ and subsequently undergoes little variation below $\Tc$ under an out-of-plane magnetic field $\mu_0 \Hout=0.1$~T [Fig.~1(b)]. The magnetization increases linearly with $\Hout$ below the saturation field $\Hc$ [Fig.~1(c)], while it shows a clear hysteresis loop under a cyclic variation in the in-plane magnetic field $\Hin$ [Fig.~1(d)]. These anisotropic responses of $M$ indicate that the interfacial ferromagnetic moments tend to align in the film plane, in contrast to the perpendicular anisotropy observed at the FeSi interfaces with insulating oxides or fluorides~\cite{Ohtsuka,Hori}. The change in the easy direction of $M$ indicates reconstruction of the {\it d}-orbital occupancy of FeSi surface state through the electronic hybridization with the Pt layer.

\begin{figure}
\begin{center}
\includegraphics*[width=8.5cm]{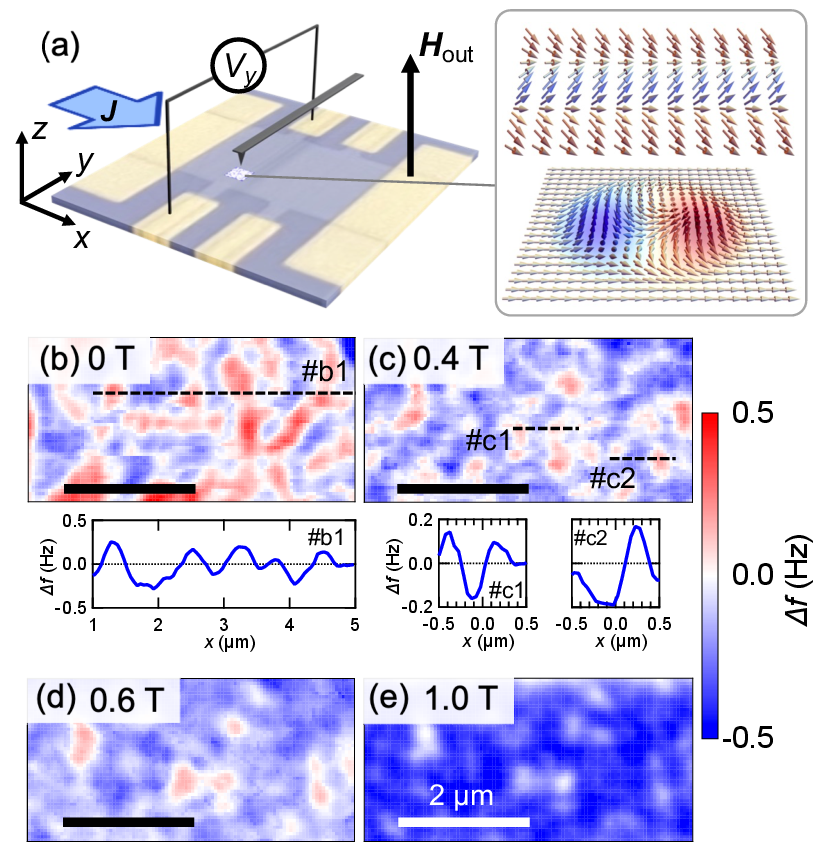}
\caption{
Magnetic force microscopy (MFM) imaging of the Hall-bar patterned Pt/FeSi bilayer under out-of-plane magnetic fields $\Hout$ at $T=10.8$~K. (a) Experimental setup for MFM imaging. The Au/Ti electrode pads (yellow regions) are used for the injection of current pulses and the detection of Hall voltages. Inset shows schematic illustrations of the expected magnetic structures, \ie, a transverse-conical state (cycloidal spin modulation with an in-plane ferromagnetic component) and a skyrmionic-bubble state (bimeron in an in-plane magnetization background). (b)-(e) MFM images at $\mu_0 \Hout= 0$~T (b), 0.4~T (c), 0.6~T (d), and 1.0~T (e). No current pulses are applied before taking the MFM images. The line scans along \#b1, \#c1, and \#c2 (dashed lines) are shown in insets of panels (b) and (c). The scale bars represent 2 $\mu$m.
}
\end{center}
\end{figure}

To further investigate the detailed magnetic structure and its responses to electric currents, we performed MFM imaging of a Hall-bar patterned Pt/FeSi under various $\Hout$ at $T=10.8$~K [Fig.~2(a)]. Here the spatial modulation of out-of-plane $M$, \ie, the local $M_z$ component, was detected as a frequency change $\Delta f$ of the vibrating cantilever magnetized along $\Hout$. 

A disordered worm-like pattern is observed at $\mu_0 \Hout =0$~T [Fig.~2(b)]. Under the application of $\mu_0 \Hout=0.4$~T, the worm-like domains are split into irregular-shaped bubble domains [Fig.~2(c)], which decrease their numbers with increasing $\Hout$ [Fig.~2(d) for  $\mu_0 \Hout=0.6$~T] and finally disappear above $\Hc$ [Fig.~2(e) for  $\mu_0 \Hout=1.0$~T]. These domains are randomly distributed and vary considerably in size and shape. The domain width is roughly 500~nm as evaluated from the line scans in Figs.~2(b) and (c). 

By analogy with the formation of spiral or skyrmionic spin textures commonly found in many heterostructures of ferromagnetic metals and heavy elements~\cite{Wiesendanger,Fert2,Everschor}, the observed submicron domains are likely to originate from the interfacial DM interaction, which should be significantly enhanced by the intrinsically strong SOC of Pt~\cite{Chshiev,Belabbes}. On the other hand, given the in-plane magnetic anisotropy in the Pt/FeSi bilayer, the magnetic domains would host different internal spin structures from the conventional cycloidal spin modulations or N\'{e}el-type skyrmions formed in the background of perpendicular magnetization. Although the in-plane spin arrangement cannot be determined in the present MFM setup, the worm-like and the bubble domains may be highly-deformed variants of transverse conical states~\cite{Yoshida} and bimerons~\cite{Kharkov,Gobel,Vaz,Ohara} [see the inset of Fig.~2(a) for their schematic illustrations]. These postulated magnetic domains are in line with the oscillation profiles observed in the line scans [Fig.~2(c), \#c1 and \#c2]. Hereafter, the isolated magnetic domains with noncollinear spin arrangements are referred to as SkBs.

\begin{figure}
\begin{center}
\includegraphics*[width=8.5cm]{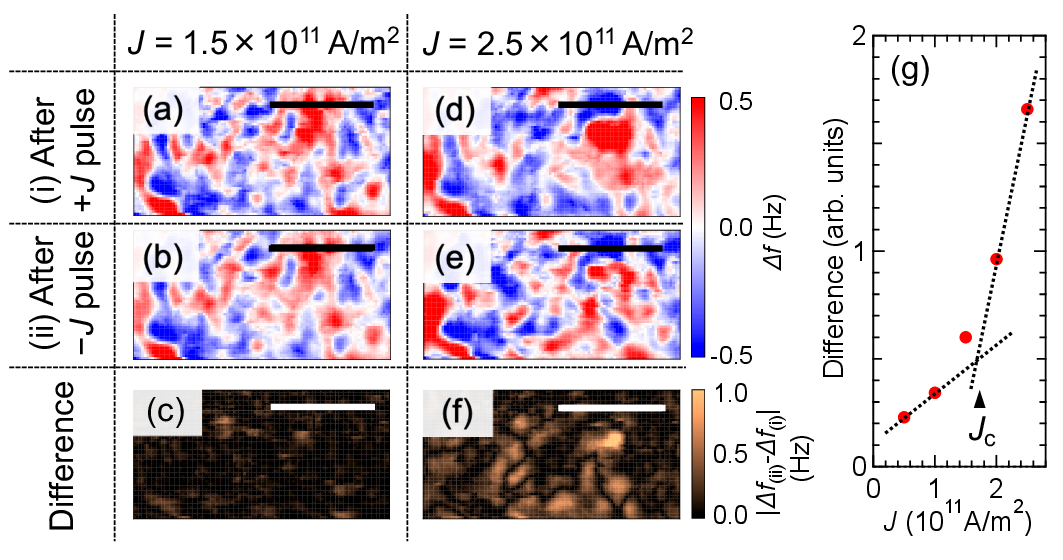}
\caption{
Current-induced changes in the magnetic domain pattern at zero magnetic field. (a) and (b) Magnetic force microscopy (MFM) images after the injection of the pulse train (\ie, successive three current pulses with current density $J=1.5\times 10^{11}$ A/m$^2$, with a duration of 10~ms, and at intervals of 1~s) in the positive (a) and negative (b) directions. (c) Difference between the MFM images [panels (a) and (b)] with the opposite current polarities. (d)-(f) MFM images [(d) and (e)] and difference image (f) in the case of the higher current density $J=2.5\times 10^{11}$ A/m$^2$. The scale bars represent 2~$\mu$m. (g) Current density $J$ dependence of the total difference integrated over the scanned area. The typical threshold current density $\Jc$ is estimated as the intersection of the dashed lines. 
}
\end{center}
\end{figure}

The substantial dispersion in size and shape of SkBs and their disordered arrangements indicate the presence of underlying structural disorders such as defects, surface roughness and grain boundaries, which produce spatial fluctuations of magnetic interactions/anisotropy and hence a nonflat energy landscape for magnetic domains~\cite{Gross,Reichhardt,Gruber}. In fact, the consequent strong pinning effects are observed as robust immobility of magnetic domains against electric currents (Fig.~3). We injected successive three current pulses with a duration of 10~ms at intervals of 1~s into the Hall bar and subsequently took a MFM image [Figs.~3(a), (b), (d), and (e)]. The current-induced changes in the magnetic-domain pattern were evaluated as the difference between MFM images with opposite current polarity [Figs.~3(c) and (f)]. Here the pulse current density $J$ is calculated assuming that the current flows homogeneously through the device [cf. Supplemental Material~\cite{supple}, Fig.~S2]. 

There is little changes in the magnetic domain pattern in response to the application of $|J|\sim 1.5 \times 10^{11}$ A/m$^2$ [Figs.~3(a)--(c)]. For the higher current pulses $|J|\sim 2.5 \times 10^{11}$ A/m$^2$, distinct transformations are spotted at various regions of the MFM image [Figs.~3(d)--(f)], while no more variations can be induced by further injection of current pulses [Supplemental Material~\cite{supple}, Fig.~S3]. 
These results represent that a fraction of magnetic domains are selectively driven perhaps in a dependent manner on the strength of the pinning force and are subsequently trapped at the strong pinning sites like grain boundaries. Interestingly, the domain pattern can be repeatedly and reproducibly alternated between the distinct trapped states [\eg, Figs.~3(d) and (e) for $J = \pm 2.5 \times 10^{11}$ A/m$^2$] by reversing the current polarity [Supplemental Material~\cite{supple}, Fig.~S3]. This may be because the strong pinning sites are closely distributed at micrometer-scale intervals and effectively constrain the random diffusion of SkBs.

The total difference, \ie, the integrated value over each difference image, shows an accelerated rise with $J$ [Fig.~3(g)], rather than an abrupt upturn across a threshold current density $J_\mathrm{c}$. This continuous change with $J$ also corroborates the local variations in the pinning potential, which lead to an expansion of the spectrum of threshold values. Nevertheless, the average $J_\mathrm{c}$ for depinning the domains can be approximated as $1.7 \times 10^{11}$ A/m$^2$ [Fig.~3(g)]. The estimated $J_\mathrm{c}$ is several orders of magnitude larger than those for conventional skyrmions~\cite{Jonietz,Yu,Jiang,Jiang2} and ranks among the highest reported values~\cite{Woo,Woo2,Hrabec}. The greatly enhanced $J_\mathrm{c}$ can be partially attributed to the reduced dimensionality of this interfacial magnetic order; this aligns with theoretical predictions that suggest an inverse relationship between $J_\mathrm{c}$ and the thickness of magnetic layer~\cite{Zang,Hoshino}.

\begin{figure}
\begin{center}
\includegraphics*[width=8.5cm]{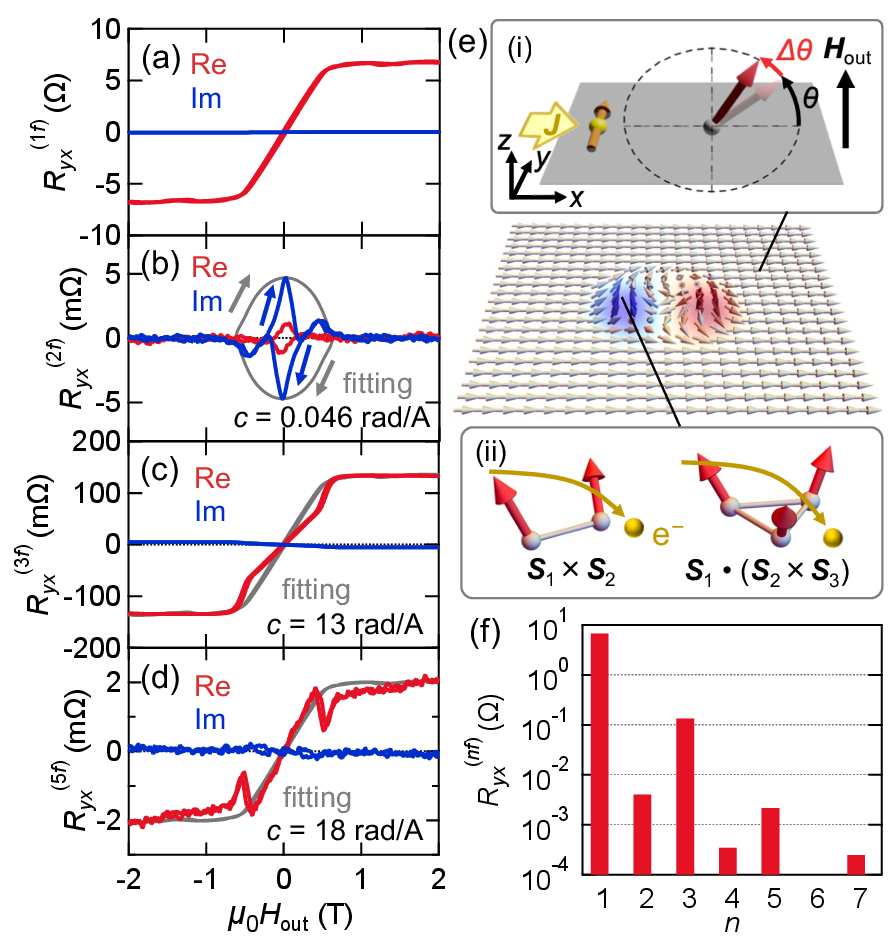}
\caption{
Fundamental $R_{yx}^{(1f)}$ and higher-harmonic $\Rnf$ Hall resistances in response to the ac current $I=I_{\mathrm{0}}\sin{(2\pi f t)}$ ($I_{\mathrm{0}}=30$ mA and $f=111$ Hz). (a)-(d) Out-of-plane magnetic-field $\Hout$ dependence of fundamental $R_{yx}^{(1f)}$ (a), second-harmonic $R_{yx}^{(2f)}$ (b), third-harmonic $R_{yx}^{(3f)}$ (c), and fifth-harmonic $R_{yx}^{(5f)}$ (d) at $T=10$~K. Fitting curves (gray lines) are produced on the basis of the spin-orbit-torque model [see main text and Supplemental Material~\cite{supple} for details]. (e) Schematic illustrations of the models for understanding mechanisms of nonlinear Hall effects: (i) oscillation of magnetization by a damping-like spin-orbit torque and (ii) deflection of conduction electrons by vector and scalar spin chiralities. (f) Comparison of the respective maximum values of $\Rnf$ ($n=1$-7) at $T=10$~K. No clear signals of forth- and sixth-harmonics are detected around zero magnetic field. [Also see Supplemental Material~\cite{supple}, Fig.~S4.]
}
\end{center}
\end{figure}

The strong pinning effects at the Pt/FeSi interface enable us to investigate Hall transport properties in the SkBs without causing their displacement or disruption under a high current below the threshold value ($I < I_\mathrm{c}$). Here, we focus on the nonlinear Hall effects, which are typically enhanced in the high-$I$ regime, in accord with the definition of nonlinear Hall resistance: $V_y =\sum_n R_{yx}^{(n)} I^n$; $R_{yx}^{(n)}$ being the $n$-th nonlinear Hall resistance. We detected $R_{yx}^{(n)}$ ($n \le 7$) by measuring higher-harmonic Hall resistances $\Rnf$ in response to an input of harmonic ac current $I=I_{\mathrm{0}}\sin{(2\pi f t)}$. 
The relationships between $\Rn$ and $\Rnf$ can be mathematically deduced in the low-frequency limit [see Supplemental Material~\cite{supple}]: the odd-order $\Rn$ is related to the real (in-phase) part of $\Rnf$ (\eg, $\mathrm{Re}[R_{yx}^{(3f)}]=-R_{yx}^{(3)} I_0^2/4$ and $\mathrm{Im}[R_{yx}^{(3f)}]=0$); the even-order $\Rn$ to the imaginary (out-of-phase) part of $\Rnf$ (\eg, $\mathrm{Re}[R_{yx}^{(2f)}]=0$ and $\mathrm{Im}[R_{yx}^{(2f)}]=-R_{yx}^{(2)} I_0/2$). 

Figures~4(a)--(d) show $\Hout$ dependence of real and imaginary parts of $\Rnf$ ($n=1$, 2, 3 and 5) at $T=10$ K under the input ac current with $I_0=30$ mA (\ie, $|J| \le 1.25 \times 10^{11}$ A/m$^{2}$)  and $f=111$ Hz. [See Supplemental Material~\cite{supple}, Fig.~S4 for other $\Rnf$.]
The linear response $R_{yx}^{1f}$ represents a conventional anomalous Hall resistance in proportion to the out-of-plane component of $M$ (\ie, $M_z$) [Fig.~4(a) and also see Fig.~1(c)]. In contrast, the higher-harmonic $\Rnf$, while being of significantly smaller magnitude as compared to $R_{yx}^{1f}$ [Fig.~4(f)], exhibit distinctive features that deviate from the proportionality with respect to $M_z$. A hysteresis loop is observed in the second-harmonic resistance, which is characterized by a pronounced peak around zero magnetic field and followed by the rapid attenuation in the fully polarized state above $\Hc$. Other even-order harmonics are not detected around zero magnetic field [Supplemental Material~\cite{supple}, Fig.~S4].
On the other hand, the third- and fifth-harmonic resistances exhibit $\Hout$-dependence profiles resembling that of $R_{yx}^{1f}$, albeit showing nonmonotonic behaviors below $\Hc$. 

As an overall trend, the second-harmonic resistance becomes pronounced below $\Hc$, while the odd-order harmonics gradually develop with increasing $H$ and reach their maxima above $\Hc$. These contrasting behaviors can be rationalized within the framework where the ac current induces oscillating motions of the background magnetization via the SOT mechanism. When the in-plane $M$ is tilted by an angle $\theta$ in the presence of $\Hout$, and this orientation is further oscillated by an angle $\Delta\theta$ due to the damping-like SOT [Fig.~4(e)], the anomalous Hall resistance can be derived as follows:
\begin{align*}
&R_{yx} = \RA M_z = \RA M \sin{(\theta + \Delta\theta)} \\
&\approx \RA M \sin{\theta} + c\RA M \cos{\theta}\cdot I -\frac{c^2}{2} \RA M \sin{\theta}\cdot I^2 +\cdots.
\end{align*}
Here we assume the linear relationship $\Delta\theta\approx c I$ in addition to the approximation by Taylor series. ($\RA$ and $c$ are coefficients.) As evident from the correspondence between this equation and the definition of $\Rn$, the even-order $\Rn$ is proportional to the in-plane component of $M$, whereas the odd-order $\Rn$ is proportional to the out-of-plane component of $M$, namely to the linear response of anomalous Hall effect $R_{yx}^{(1)}$. This model effectively captures the general behaviors of all higher-harmonic resistances [see gray fitting curves in Figs.~4(b)--(d) and also see Supplemental Material~\cite{supple} for the detailed fitting curves], despite discernible variation in $c$ to fit $\Rnf$ curves, particularly the reduced $c$ for reproducing $R_{yx}^{(2f)}$ [Supplemental Material~\cite{supple}, Fig.~S4]. This modest discrepancy in the fitting coefficients for the odd-order $\Rnf$ suggests the necessity to extend the relationship between $\Delta\theta$ and $I$ to higher orders (\ie, $\Delta\theta \approx cI + c'I^2 + c'' I^3+\cdots$) and improve the approximation accuracy. In contrast, the reduced $c$ for $R_{yx}^{(2f)}$ may be ascribed to the existence of multiple domains with different in-plane magnetization directions. There occurs strong cancellation between the contributions from the domains with oppositely-polarized background magnetization. Only an uncompensated $R_{yx}^{(2f)}$, which appears as the reduction of $c$, survives as a result of an accidental imbalance of magnetic domain population. The same thing happens to other even-order $\Rnf$, which are undetectably small [Fig.~4(f) and Supplemental Material~\cite{supple}, Fig.~S4].

The origin of the nonmonotonic behaviors, which cannot be explained by the above model, may be related to the noncollinear texture of the SkBs, as these characteristic features only show up below $\Hc$. One possible microscopic origin could be asymmetric scatterings of the conduction electrons caused by the current-induced excitations of the noncollinear spin structures~\cite{Ishizuka,Ishizuka2,Schutte}. On the basis of symmetry argument, the higher-order nonlinear Hall effects are indeed expressed using vector and/or scalar spin chiralities [Fig.~4(e)] as descriptors of the noncollinearity. Given the symmetry of the heterochiral FeSi thin film (\ie, $D_3$ symmetry) and SO(3) spin rotational symmetry, the nonlinear Hall effects bear proportionate relationships to the scalar spin chirality, as is the linear response of topological Hall effect: $J_y^{(n)} \propto \int{\bm{S}\cdot (\partial_x \bm{S} \times \partial_y \bm{S})} \mathrm{d}x \mathrm{d}y \cdot E_x^{n}$, where $\bm{S}(\bm{r})$ represents a spin texture. On the other hand, the nonmonotonic profiles of $\Rnf$ show different $\Hout$ dependences with respect to the order $n$: the sharp anomaly appears around $\Hc$ in the $R_{yx}^{(5f)}$ profile; while the broad dip structures are discerned in the middle of the noncollinear-spin phase below $\Hc$ as for $R_{yx}^{(2f)}$ and $R_{yx}^{(3f)}$ profiles. This indicates that other mechanisms possibly exist for forming the complex behaviors of nonlinear Hall effects arising from the spin-noncollinearity. For example, enhanced emergent electrodynamics upon the collapse of SkBs~\cite{Milde,Schutte2,Kanazawa} may explain the anomaly around $\Hc$ observed in $R_{yx}^{(5f)}$.

\section{Conclusion}
We have realized the formation of SkBs at the quasi-2D ferromagnetic surface state of FeSi by leveraging the SOC proximity of the adjacent Pt layer. The SkBs are strongly pinned by the structural disorders, resulting in their polydispersity/polymorphism and the greatly enhanced depinning threshold $\Jc\approx 1.7\times 10^{11}$ A/m$^2$. The strong pinning effect preserves the assembled structure of SkBs even at high current densities below $\Jc$, enabling the access to the regime where prominent nonlinear responses show up. Consequently, we have identified the higher-order nonlinear Hall resistances $\Rn$ ($2 \le n \le 7$) with the characteristic $H$ dependences, which comprise (i) the primary contribution from the magnetization dynamics due to SOT and (ii) the nontrivial Hall effects related to the noncollinearity inherent in the topological spin textures. Notably, the present discoveries of the strongly pinned SkBs and their nonlinear responses may find application in the emerging field of reservoir computing~\cite{Prychynenko,Pinna,Yokouchi2,Raab,OLee,Grollier,Everschor2}. While van der Waals heterostructures have enabled the tailored manipulation of various quantum functionalities, a similar potential may exist in designed interfacial states of FeSi.

\section{Acknowledgments}
We thank X. Z. Yu and S. Mori for experimental supports and J. Masell for fruitful discussions. This work was supported by JSPS KAKENHI (Grants No. 22K18965, No. 23H04017, No. 23H05431 and No. 23H05462), JST FOREST (Grant No. JPMJFR2038), JST CREST (Grant No. JPMJCR1874 and No. JPMJCR23O3), and Mitsubishi Foundation.


\begin{thebibliography}{100}
\bibitem{Fert} M. N. Baibich, J. M. Broto, A. Fert, F. Nguyen Van Dau, F. Petroff, P. Etienne, G. Creuzet, A. Friederich, and J. Chazelas, Phys. Rev. Lett. {\bf 61}, 2472 (1988).
\bibitem{Grunberg} G. Binasch, P. Gr\"{u}nberg, F. Saurenbach, and W. Zinn, Phys. Rev. B {\bf 39}, 4828(R) (1989).
\bibitem{Parkin} S. S. P. Parkin, Annu. Rev. Mater. Sci. {\bf 25}, 357-388 (1995).
\bibitem{Cortie} D. L. Cortie, G. L. Causer, K. C. Rule, H. Fritzsche, W. Kreuzpaintner, and F. Klose, Adv. Funct. Mater. {\bf 30}, 1901414 (2020).
\bibitem{Tokura} Y. Tokura, K. Yasuda, and A. Tsukazaki, Nat. Rev. Phys. {\bf 1}, 126-143 (2019).
\bibitem{Bernevig} B. A. Bernevig, C. Felser, and H. Beidenkopf, Nature {\bf 603}, 41-51 (2022).
\bibitem{Smejkal} L. \v{S}mejkal, Y. Mokrousov, B. Yan, and A. H. MacDonald, Nat. Phys. {\bf 14}, 242-251 (2014).
\bibitem{Hasan} M. Z. Hasan, G. Chang, I. Belopolski, G. Bian, S.-Y. Xu, and J.-X. Yin, Nat. Rev. Mater. {\bf 6}, 784-803 (2021).
\bibitem{Geim} A. K. Geim and I. V. Grigorieva, Nature {\bf 499}, 419-425 (2013).
\bibitem{Gong} C. Gong and X. Zhang, Science {\bf 363}, eaav4450 (2019).
\bibitem{Gibertini} M. Gibertini, M. Koperski, A. F. Morpurgo, and K. S. Novoselov, Nat. Nanotechnol. {\bf 14}, 408-419 (2019).
\bibitem{Mak} K. F. Mak, J. Shan, and D. C. Ralph, Nat. Rev. Phys. {\bf 1}, 646-661 (2019).
\bibitem{Xu} B. Huang, M. A. McGuire, A. F. May, D. Xiao, P. Jarillo-Herrero, and X. Xu, Nat. Mater. {\bf 19}, 1276-1289 (2020).
\bibitem{Chang} C.-Z. Chang, J. Zhang, X. Feng, J. Shen, Z. Zhang, M. Guo, K. Li, Y. Ou, P. Wei, L.-L. Wang, Z.-Q. Ji, Y. Feng, S. Ji, X. Chen, J. Jia, X. Dai, Z. Fang, S.-C. Zhang, K. He, Y. Wang, L. Lu, X.-C. Ma, and Q.-K. Xue, Science {\bf 340}, 167-170 (2013).
\bibitem{Deng} Y. Deng, Y. Yu, M. Z. Shi, Z. Guo, Z. Xu, J. Wang, X. H. Chen, and Y. Zhang, Science {\bf 367}, 895-900 (2020).
\bibitem{Serlin} M. Serlin, C. L. Tschirhart, H. Polshyn, Y. Zhang, J. Zhu, K. Watanabe, T. Taniguchi, L. Balents, and A. F. Young, Science {\bf 367} 900-903 (2020).
\bibitem{Park} H. Park, J. Cai, E. Anderson, Y. Zhang, J. Zhu, X. Liu, C. Wang, W. Holtzmann, C. Hu, Z. Liu, T. Taniguchi, K. Watanabe, J.-H. Chu, T. Cao, L. Fu, W. Yao, C.-Z. Chang, D. Cobden, D. Xiao, and X. Xu, Nature {\bf 622}, 74-79 (2023).
\bibitem{Zhong} D. Zhong, K. L. Seyler, X. Linpeng, R. Cheng, N. Sivadas, B. Huang, E. Schmidgall, T. Taniguchi, K. Watanabe, M. A. McGuire, W. Yao, D. Xiao, K.-M. C. Fu, X. Xu , Sci. Adv. {\bf 3}, e1603113 (2017).
\bibitem{Seyler} K. L. Seyler, D. Zhong, B. Huang, X. Linpeng, N. P. Wilson, T. Taniguchi, K. Watanabe, W. Yao, D. Xiao, M. A. McGuire, K.-M. C. Fu, and X. Xu, Nano Lett. {\bf 18}, 3823-3828 (2018).
\bibitem{Manchon} A. Manchon, H. C. Koo, J. Nitta, S. M. Frolov, and R. A. Duine, Nat. Mater. {\bf 14}, 871-882 (2015).
\bibitem{Panagopoulos}A. Soumyanarayanan, N. Reyren, A. Fert, and C. Panagopoulos, Nature {\bf 539}, 509-517 (2016).
\bibitem{Hellman} F. Hellman, A. Hoffmann, Y. Tserkovnyak, G. Beach, E. Fullerton, C. Leighton, A. MacDonald, D. Ralph, D. Arena, H. Durr, P. Fischer, J. Grollier, J. Heremans, T. Jungwirth, A. Kimmel, B. Koopmans, I. Krivorotov, S. May, A. Petford-Long, J. Rondinelli, N. Samarth, I. Schuller, A. Slavin, M. Stiles, O. Tchernyshyov, A. Thiaville, and B. Zink, Rev. Mod. Phys. {\bf 89}, 025006 (2017).
\bibitem{Bihlmayer} G. Bihlmayer, P. No\"{e}l, D. V. Vyalikh, E. V. Chulkov, and A. Manchon, Nat. Rev. Phys. {\bf 4}, 642-659 (2022).
\bibitem{Rashba} E. I. Rashba, Sov. Phys. Solid State {\bf 2}, 1109-1122 (1960). 
\bibitem{Dzyaloshinsky} I. Dzyaloshinsky, J. Phys. Chem. Solids {\bf 4}, 241-255 (1958).
\bibitem{Moriya} T. Moriya, Phys. Rev. {\bf 120}, 91-98 (1960).
\bibitem{Wiesendanger} R. Wiesendanger, Nat. Rev. Mater. {\bf 1}, 16044 (2016).
\bibitem{Fert2} A. Fert, N. Reyren, and V. Cros, Nat. Rev. Mater. {\bf 2}, 17031 (2017).
\bibitem{Bogdanov} A. N. Bogdanov and C. Panagopoulos, Nat. Rev. Phys. {\bf 2}, 492-498 (2020).
\bibitem{Edelstein} V. M. Edelstein, Solid State Commun. {\bf 73}, 233-235 (1990).
\bibitem{Garello}K. Garello, I. M. Miron, C. O. Avci, F. Freimuth, Y. Mokrousov, S. Bl\"{u}gel, S. Auffret, O. Boulle, G. Gaudin, and P. Gambardella, Nature {\bf 465}, 901-904 (2010).
\bibitem{Manchon2} A. Manchon, J. \v{Z}elezn\'{y}, I. M. Miron, T. Jungwirth, J. Sinova, A. Thiaville, K. Garello, and P. Gambardella, Rev. Mod. Phys. {\bf 91}, 035004 (2019).
\bibitem{Yokouchi} T. Yokouchi, N. Kanazawa, A. Kikkawa, D. Morikawa, K. Shibata, T. Arima, Y. Taguchi, F. Kagawa, and Y. Tokura, Nat. Commun. {\bf 8}, 866 (2017).
\bibitem{Tokura2} Y. Tokura and N. Nagaosa, Nat. Commun. {\bf 9}, 3740 (2018).
\bibitem{Neubauer} A. Neubauer, C. Pfleiderer, B. Binz, A. Rosch, R. Ritz, P. G. Niklowitz, and P. B\"{o}ni, Phys. Rev. Lett. {\bf 102}, 186602 (2009).
\bibitem{Lee} Minhyea Lee, W. Kang, Y. Onose, Y. Tokura, and N. P. Ong, Phys. Rev. Lett. {\bf 102}, 186601 (2009).
\bibitem{Jonietz} F. Jonietz, S. M\"{u}hlbauer, C. Pfleiderer, A. Neubauer, W. M\"{u}nzer, A. Bauer, T. Adams, R. Georgii, P. B\"{o}ni, R. A. Duine, K. Everschor, M. Garst, and A. Rosch, Science {\bf 330}, 1648-1651 (2010).
\bibitem{Woo} S. Woo, K. Litzius, B. Kr\"{u}ger, M.-Y. Im, L. Caretta, K. Richter, M. Mann, A. Krone, R. M. Reeve, M. Weigand, P. Agrawal, I. Lemesh, M.-A. Mawass, P. Fischer, M. Kl\"{a}ui, and G. S. D. Beach, Nat. Mater. {\bf 15}, 501-506 (2016).
\bibitem{Ohtsuka} Y. Ohtsuka, N. Kanazawa, M. Hirayama, A. Matsui, T. Nomoto, R. Arita, T. Nakajima, T. Hanashima, V. Ukleev, H. Aoki, M. Mogi, A. Tsukazaki, M. Ichikawa, M. Kawasaki, and Y. Tokura, Sci. Adv. {\bf 7}, eabj0498 (2021).
\bibitem{Hori} T. Hori, N. Kanazawa, M. Hirayama, K. Fujiwara, A. Tsukazaki, M. Ichikawa, M. Kawasaki, and Y. Tokura, Adv. Mater. {\bf 35}, 2206801 (2023).
\bibitem{Maple} Y. Fang, S. Ran, W. Xie, S. Wang, Y. S. Meng, M. B. Maple, Proc. Natl. Acad. Sci. U. S. A. {\bf 115}, 8558 (2018).
\bibitem{Barth} B. Yang, M. Uphoff, Y.-Q. Zhang, J. Reichert, A. P. Seitsonen,
A. Bauer, C. Pfleiderer, J. V. Barth, Proc. Natl. Acad. Sci. U. S. A. {\bf 118}, e2021203118 (2021).
\bibitem{Paglione} Y. S. Eo, K. Avers, J. A. Horn, H. Yoon, S. R. Saha, A. Suarez, M. S. Fuhrer, and J. Paglione, Appl. Phys. Lett. {\bf 122}, 233102 (2023).
\bibitem{Mattheiss} L. F. Mattheiss and D. R. Hamann, Phys. Rev. B {\bf 47}, 13114-13119 (1993).
\bibitem{Zak} J. Zak, Phys. Rev. Lett. {\bf 62}, 2747-2750 (1989).
\bibitem{Resta} R. Resta, Ferroelectrics {\bf 136}, 51-55 (1992).
\bibitem{King} R. D. King-Smith and D. Vanderbilt, Phys. Rev. B {\bf 47}, 1651-1654 (1993).
\bibitem{Vanderbilt} D. Vanderbilt and R. D. King-Smith, Phys. Rev. B {\bf 47}, 4442-4455 (1993).
\bibitem{Changdar} S. Changdar, S. Aswartham, A. Bose, Y. Kushnirenko, G. Shipunov, N. C. Plumb, M. Shi, A. Narayan, B. B\"{u}chner, S. Thirupathaiah, Phys. Rev. B {\bf 101}, 235105 (2020).
\bibitem{Aeppli} Z. Schlesinger, Z. Fisk, H.-T. Zhang, M. B. Maple, J. DiTusa, and G. Aeppli, Phys. Rev. Lett. {\bf 71}, 1748-1751 (1993).
\bibitem{Doniach} C. Fu and S. Doniach, Phys. Rev. B {\bf 51}, 17439-17445 (1995).
\bibitem{Anisimov} V. I. Anisimov, S. Y. Ezhov, I. S. Elfimov, I. V. Solovyev, and T. M. Rice, Phys. Rev. Lett. {\bf 76}, 1735-1738 (1996).
\bibitem{Kotliar} J. M. Tomczak, K. Haule, and G. Kotliar, Proc. Natl. Acad. Sci. U.S.A. {\bf 109}, 3243-3246 (2012).
\bibitem{Everschor} K. Everschor-Sitte, J. Masell, R. M. Reeve, and M. Kl\"{a}ui, J. Appl. Phys. {\bf 124}, 240901 (2018).
\bibitem{Sirringhaus} H. von K\"{a}nel, K. A. M\"{a}der, E. M\"{u}ller, N. Onda, and H. Sirringhaus, Phys. Rev. B {\bf 45}, 13807(R) (1992).
\bibitem{Chshiev} H. Yang, A. Thiaville, S. Rohart, A. Fert, and M. Chshiev, Phys. Rev. Lett. {\bf 115} 267210 (2015).
\bibitem{Belabbes} A. Belabbes, G. Bihlmayer, F. Bechstedt, S. Bl\"{u}gel, and A. Manchon, Phys. Rev. Lett. {\bf 117} 247202 (2016).
\bibitem{Yoshida} Y. Yoshida, S. Schr\"{o}der, P. Ferriani, D. Serrate, A. Kubetzka, K. von Bergmann, S. Heinze, and R. Wiesendanger, Phys. Rev. Lett. {\bf 108}, 087205 (2012).
\bibitem{Kharkov} Y. A. Kharkov, O. P. Sushkov, and M. Mostovoy, Phys. Rev. Lett. {\bf 119}, 207201 (2017).
\bibitem{Gobel} B. G\"{o}bel, A. Mook, J. Henk, I. Mertig, and O. A. Tretiakov, Phys. Rev. B {\bf 99}, 060407(R) (2019).
\bibitem{Vaz} J. Vijayakumar, Y. Li, D. Bracher, C. W. Barton, M. Horisberger, T. Thomson, J. Miles, C. Moutafis, F. Nolting, and C.A.F. Vaz, Phys. Rev. Applied {\bf 14}, 054031 (2020).
\bibitem{Ohara} K. Ohara, X. Zhang, Y. Chen, S. Kato, J. Xia, M. Ezawa, O. A. Tretiakov, Z. Hou, Y. Zhou, G. Zhao, J. Yang, and X. Liu, Nano Lett. {\bf 22}, 8559-8566 (2022).
\bibitem{Gross} I. Gross, W. Akhtar, A. Hrabec, J. Sampaio, L. J. Mart\'{i}nez, S. Chouaieb, B. J. Shields, P. Maletinsky, A. Thiaville, S. Rohart, and V. Jacques, Phys. Rev. Materials {\bf 2}, 024406 (2018).
\bibitem{Reichhardt} C. Reichhardt, C. J. O. Reichhardt, and M. V. Milo\v{s}evi\'{c}, Rev. Mod. Phys. {\bf 94}, 035005 (2022).
\bibitem{Gruber} R. Gruber, J. Z\'{a}zvorka, M. A. Brems, D. R. Rodrigues, T. Dohi, N. Kerber, B. Seng, M. Vafaee, K. Everschor-Sitte, P. Virnau, and M. Kl\"{a}ui, Nat. Commun. {\bf 13}, 3144 (2022).
\bibitem{supple} See Supplemental Material at [url] for characterization of the sample, longitudinal resistance, current-pulse dependence of the development of magnetic-domain pattern, and dataset of all higher-harmonic Hall resistances.
\bibitem{Yu} X.Z. Yu, N. Kanazawa, W.Z. Zhang, T. Nagai, T. Hara, K. Kimoto, Y. Matsui, Y. Onose, and Y. Tokura, Nat. Commun. {\bf 3}, 988 (2012).
\bibitem{Jiang} W. Jiang, P. Upadhyaya, W. Zhang, G. Yu, M. B. Jungfleisch, F. Y. Fradin, J. E. Pearson, Y. Tserkovnyak, K. L. Wang, O. Heinonen, S. G. E. te Velthuis, A. Hoffmann, Science {\bf 349} 283-286 (2015).
\bibitem{Jiang2} W. Jiang, X. Zhang, G. Yu, W. Zhang, X. Wang, M. B. Jungfleisch, J. E. Pearson, X. Cheng, O. Heinonen, K. L. Wang, Y. Zhou, A. Hoffmann, and S. G. E. te Velthuis, Nat. Phys. {\bf 13}, 162-169 (2017). 
\bibitem{Hrabec} A. Hrabec, J. Sampaio, M. Belmeguenai, I. Gross, R. Weil, S. M. Ch\'{e}rif, A. Stashkevich, V. Jacques, A. Thiaville, and S. Rohart, Nat. Commun. {\bf 8}, 15765 (2017).
\bibitem{Woo2} S. Woo, K. M. Song, X. Zhang, Y. Zhou, M. Ezawa, X. Liu, S. Finizio, J. Raabe, N. J. Lee, S. Kim, S.-Y. Park, Y. Kim, J.-Y. Kim, D. Lee, O. Lee, J. W. Choi, B.-C. Min, H. C. Koo, and J. Chang, Nat. Commun. 9, 959 (2018). 
\bibitem{Zang} J. Zang, M. Mostovoy, J. H. Han, and N. Nagaosa, Phys. Rev. Lett. {\bf 107}, 136804 (2011).
\bibitem{Hoshino} S. Hoshino and N. Nagaosa, Phys. Rev. B {\bf 97}, 024413 (2018).
\bibitem{Ishizuka}H. Ishizuka and N. Nagaosa, Sci. Adv. {\bf 4}, eaap9962 (2018).
\bibitem{Ishizuka2} H. Ishizuka and N. Nagaosa, Nat. Commun. {\bf 11}, 2986 (2020).
\bibitem{Schutte} C. Sch\"{u}tte and M. Garst, Phys. Rev. B {\bf 90}, 094423 (2014).
\bibitem{Milde} P. Milde, D. K\"{o}hler, J. Seidel, L. M. Eng, A. Bauer, A. Chacon, J. Kindervater, S. M\"{u}hlbauer, C. Pfleiderer, S. Buhrandt, C. Sch\"{u}tte, and A. Rosch, Science {\bf 340}, 1076-1080 (2013).
\bibitem{Schutte2} C. Sch\"{u}tte and A. Rosch, Phys. Rev. B {\bf 90} 174432 (2014).
\bibitem{Kanazawa} N. Kanazawa, Y. Nii, X. -X. Zhang, A. S. Mishchenko, G. De Filippis, F. Kagawa, Y. Iwasa, N. Nagaosa, and Y. Tokura, Nat. Commun. {\bf 7}, 11622 (2016). 
\bibitem{Prychynenko} D. Prychynenko, M. Sitte, K. Litzius, B. Kr\"{u}ger, G. Bourianoff, M. Kl\"{a}ui, J. Sinova, and K. Everschor-Sitte, Phys. Rev. Applied {\bf 9}, 014034 (2018).
\bibitem{Pinna} D. Pinna, G. Bourianoff, and K. Everschor-Sitte, Phys. Rev. Applied {\bf 14}, 054020 (2020).
\bibitem{Yokouchi2} T. Yokouchi, S. Sugimoto, B. Rana, S. Seki, N. Ogawa, Y. Shiomi, S. Kasai, and Y. Otani, Sci. Adv. {\bf 8}, eabq5652 (2022).
\bibitem{Raab} K. Raab, M. A. Brems, G. Beneke, T. Dohi, J. Roth\"{o}rl, F. Kammerbauer, J. H. Mentink, M. Kl\"{a}ui, Nat. Commun. {\bf 13}, 6982 (2022).
\bibitem{OLee} Oscar Lee, Tianyi Wei, Kilian D Stenning, Jack C Gartside, Dan Prestwood, Shinichiro Seki, Aisha Aqeel, Kosuke Karube, Naoya Kanazawa, Yasujiro Taguchi, Christian Back, Yoshinori Tokura, Will R Branford, and Hidekazu Kurebayashi, arXiv:2209.06962 (2022).
\bibitem{Grollier} J. Grollier, D. Querlioz, K. Y. Camsari, K. Everschor-Sitte, S. Fukami, and M. D. Stiles, Nat. Electron. {\bf 3}, 360-370 (2020). 
\bibitem{Everschor2} O. Lee, R. Msiska, M. A. Brems, M. Kl\"{a}ui, H. Kurebayashi, K. Everschor-Sitte, Appl. Phys. Lett. {\bf 122}, 260501 (2023).

\end{thebibliography}
\end{document}